\documentclass[sigconf]{acmart}
\usepackage{array}
\usepackage{graphicx}
\usepackage{clrscode}
\usepackage{balance}
\usepackage{subfigure}
\usepackage{multirow}
\usepackage{float}
\usepackage{color}
\usepackage{soul}
\usepackage{mdframed}
\usepackage{pmat}
\usepackage{amsopn}
\usepackage{mathrsfs}
\usepackage{mathtools}
\usepackage{amsmath}
\usepackage{arydshln}
\usepackage{hyperref}
\usepackage{multicol}
\usepackage{blkarray}
\usepackage{enumerate}
\usepackage{courier}
\usepackage{rotating}
\usepackage{booktabs}
\usepackage{diagbox}
\usepackage{makecell}
\usepackage[lined,boxruled,commentsnumbered,linesnumbered]{algorithm2e}

\setcopyright{rightsretained}
\acmConference[EARS'19]{SIGIR 2019 Workshop on ExplainAble Recommendation and Search}{July 25, 2019}{Paris, France}
\acmDOI{10.475/123_4}
\acmISBN{123-4567-24-567/08/06}
\acmYear{2019}
\copyrightyear{2019}

\begin{document}

%\title{Abstractive Natural Language Explanations Generation for Personalized Recommendation}
%\title{Generating Natural Language Explanations for \\Personalized Recommendation}
\title{Generate Natural Language Explanations for Recommendation}
%\title{Explainable Recommendation based on Natural Language Generation}

%\author{Hanxiong Chen$^\dagger$, Xu Chen$^\ddagger$, Shaoyun Shi$^\ddagger$, Yongfeng Zhang$^\dagger$}
%\affiliation{\institution{$^\dagger$Rutgers University, $^\ddagger$Tsinghua University}}
%\email{{hanxiong.chen, yongfeng.zhang}@rutgers.edu, {xu-ch14,shisy17}@mails.tsinghua.edu.cn}

\author{Hanxiong Chen}
\affiliation{\institution{Rutgers University}}
\email{hanxiong.chen@rutgers.edu}

\author{Xu Chen}
\affiliation{\institution{Tsinghua University}}
\email{xu-ch14@mails.tsinghua.edu.cn}

\author{Shaoyun Shi}
\affiliation{\institution{Tsinghua University}}
\email{shisy17@mails.tsinghua.edu.cn}

\author{Yongfeng Zhang}
\affiliation{\institution{Rutgers University}}
\email{yongfeng.zhang@rutgers.edu}

%%%%%%%%%%%%%%%%%%%%%%%%%%%%%%%%%%%%%%%%%%%%%%%%
\begin{abstract}
Providing personalized explanations for recommendations can help users to understand the underlying insight of the recommendation results, which is helpful to the effectiveness, transparency, persuasiveness and trustworthiness of recommender systems. Current explainable recommendation models mostly generate textual explanations based on pre-defined sentence templates. However, the expressiveness power of template-based explanation sentences are limited to the pre-defined expressions, and manually defining the expressions require significant human efforts.

%Motivated by this problem, we propose to develop a deep neural network model called \textbf{PRE} which can not only automatically generate natural language explanations to inform users the reasons of the recommended result, but also make precise rating prediction in the meanwhile. For text explanation generation part, we design a word-level hierarchical gated recurrent unit (GRU) with a combination of attention mechanism to generate personalized explanation sentences. To improve the training data quality, we make use of external knowledge --- feature words --- to filter out useless review sentences for denoising. Since both text generation and rating prediction model use the same user and item latent representations as input, our experiments show that the rating quality would be significantly improved by learning from text information. Moreover, the generated explanation sentences are human readable and cover most of the feature words of a real user review. 

Motivated by this problem, we propose to generate free-text natural language explanations for personalized recommendation. In particular, we propose a hierarchical sequence-to-sequence model (HSS) for personalized explanation generation. Different from conventional sentence generation in NLP research, a great challenge of explanation generation in e-commerce recommendation is that not all sentences in user reviews are of explanation purpose. To solve the problem, we further propose an auto-denoising mechanism based on topical item feature words for sentence generation. Experiments on various e-commerce product domains show that our approach can not only improve the recommendation accuracy, but also the explanation quality in terms of the offline measures and feature words coverage. This research is one of the initial steps to grant intelligent agents with the ability to explain itself based on natural language sentences. 

%both offline measures and online user judgements
\end{abstract}




\maketitle

%%%%%%%%%%%%%%%%%%%%%%%%%%%%%%%%%%%%%%%%%%%%%%%%%%%
\section{Introduction}
Recommender systems are playing an important role in many online applications. They provide personalized suggestions to help user select the most relevant items based on their preferences. Collaborative Filtering (CF) has been one of the most successful approaches to generate recommendations based on historical user behaviors \cite{ricci2015recommender}. However, the recently popular latent representation approaches to CF -- including both shallow or deep models -- can hardly explain their rating prediction and recommendation results to users \cite{zhang2018explainable}.
%the numerical rating score itself lacks of explainability and users are not able to intuitively understand the reasons why the products are recommended. 

Researchers have noticed that appropriate explanations are important to recommendation systems
%The importance and necessity of explainability in recommender system was already indicated by Herlocker et al. in 2000
\cite{zhang2018explainable}, which can help to improve the system effectiveness, transparency, persuasiveness and trustworthiness. As a result, researchers have looked into explainable recommendation \cite{zhang2018explainable} and search \cite{ai2019explainable} in the recent years \cite{zhang2014explicit,zhang2018explainable,ren2017social,chen2018neural,li2017neural,chen2016learning,wang2018explainable,xian2020cafe,xian2019reinforcement,ai2018learning,chen2019personalized,chen2019dynamic,li2020generate},
%which shows that this is a significant and valuable research work. 
%On the other hand, to make the recommended results explainable is also following the urgent requirements of explainable artificial intelligence (AI). The new European General Data Protection Regulation (GDPR and ISO/IEC 27001) entering into force recently, which requires the AI models must have a possibility to make the results re-traceable on demand \cite{DBLP:journals/corr/abs-1712-09923}. We can anticipate that similar regulations, which restrict the business use of black-box AI approaches, will not take a long time to be announced in other countries. Therefore, we propose this idea to generate personalized natural language explanations for recommender systems. This work can not only improve the performance of recommender systems but also reflect our foresight to the future situation. 
which can not only provide users with the recommendation lists, but also intutive explanations about why these items are recommended. 
%In this way, it helps to improve the effectiveness, persuasiveness, trustworthiness and user satisfaction of recommender systems \cite{zhang2018explainable}. 

Recommendation explanations can be provided in many different forms, and among the many, a frequently used one is textual sentence explanation. 
%way to do an explainable recommendation is to present textual explanation sentences.
Current textual explanation generation models can be broadly classified into two categories -- template-based methods and retrieval-based methods. Template-based models, such as \cite{zhang2014explicit,wang2018explainable}, define one or more explanation sentence templates, and then fill different words into the templates according to the corresponding recommendation so as to generate different explanations. Such words could be, for example, item feature words that the target user is interested in. However, template-based method requires extensive human efforts to define different templates for different scenarios,
%to define multiple templates for various scenarios. 
and it limits the expressive power of explanation sentences to the pre-defined templates.
%try to do textual analysis on user review data to extract feature words according to users preferences. Then these words would be filled into predefined templates to generate an explanation. 
Retrieval-based methods such as \cite{chen2018neural}, on the other hand, attempt to retrieve particular sentences from user reviews as the explanations of a recommendation, which improves the expression diversity of explanation sentences. 
However, the explanations are limited to existing sentences and the model cannot produce new sentences for explanation. 

\begin{figure}[t]
\centering
\includegraphics*[scale=0.345]{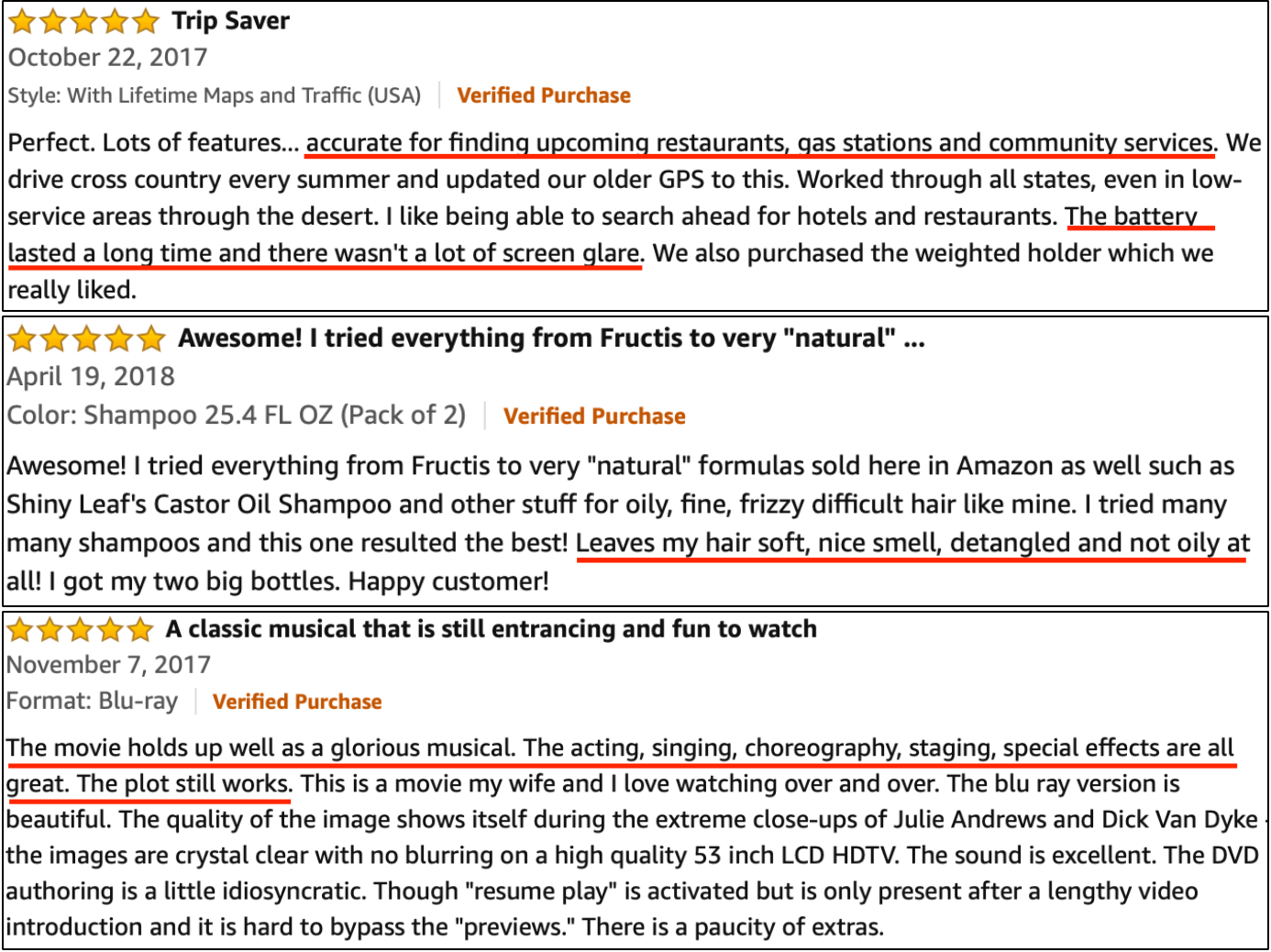}
\vspace{-15pt}
\caption{An example of user reviews in e-commerce. The sentences with red underlines are good for explanations.}
\vspace{-15pt}
\label{fig:example}
\end{figure}

Considering these problems, we propose to conduct explainable recommendation by generating free-text natural language explanations, meanwhile keep a high prediction accuracy. 
%a novel neural network framework which can simultaneously do personalized abstractive natural language explanation generation and make precise rating prediction. The abstractive natural language explanation stands for non retrieval-based model and no predefined templates are needed. 
There exist three key challenges to build and evaluate a personalized natural language explanation system. 1) Data bias -- the most commonly used text resources for training explainable recommender systems are user-generated reviews. 
%Currently, we do not have a good dataset especially for explanations. 
Although the reviews are plentiful, informative and contain valuable information about users opinions and product features \cite{zhang2014explicit,lu2018coevolutionary}, they can be very noisy and not all the sentences in a review are of explanation purpose. Take Figure \ref{fig:example} as an example, only the underlined sentence is really commenting about the product. To train a good explanation generator, our model should have the ability of auto-denoising so as to focus on the training of explanation sentences. 2) Personalization -- since different users may pay attention to different product features, a good explainable recommendation system should have the ability to provide tailored explanations for different users according to the features that the user cares about. 3) Evaluation -- although explainable recommendation has been widely researched in recent years, our understanding is still limited regarding which metric(s) is appropriate to evaluate the explainability of explanations. Recent research adopt readability measures in NLP (such as ROUGE score) for evaluation, but since explainability is not equivalent to readability, only generating readable sentences is not sufficient, and we need to take the effectiveness of recommendation into consideration. This problem involves deep understandings of natural language, and it also contributes technical merit to natural language processing research.

Motivated by these challenges, we propose a hierarchical sequence-to-sequence model (HSS) with auto-denoising for personalized recommendation and natural language explanation generation.
%we propose to design a novel deep neural network framework named \textbf{PRE} to do abstractive natural language explanations generation for personalized recommendation and our extensive experiments also prove that our work can significantly improve the rating quality. 
In particular, the paper makes the following contributions:
%\begin{itemize}

$\bullet$ We design a hierarchical generation model, which is able to collaboratively learn over multiple sentences from different users for explanation sentence generation.
%combine with the feature words extracted by toolkit [cite?], to filter out noise sentences to improve the training quality.

$\bullet$ Based on item feature words extracted from reviews, we design a feature-aware attention model to implicitly select explanation sentences from reviews for model learning,
%We also borrow the idea of \cite{xing2017topic} to 
and we further introduce a feature attention model to enhance the feature-level personality of the explanations. 
%Inspired by paper \cite{tang2016context}, we add user and item embeddings into text generation process to improve the personality.

$\bullet$ We adopt three offline metrics -- BLEU score, ROUGE score and feature coverage -- to evaluate the quality of the generated explanations. The first two metrics are classical measures for neural machine translation and text summarization. BLEU score is precision-based while ROUGE score is relatively recall-based. They are complement to each other and it would be reasonable to report both scores to reflect the quality of machine generated text. We also use feature words coverage to show how well a model can capture the real user personalized preferences. In the meanwhile, the feature words coverage is a possible measure of the explainability of the generated explanation sentences. 
%besides, we further conduct online user experiments to analyze the correlation between explainability and these metrics. In this way, we make a first step to explore how the frequently used sentence evaluation metrics approximate explainability in recommender systems.
%analysis on both online and offline results to figure out which offline metric would be better for evaluating explainability.
%\end{itemize}
%\cite{lin2004rouge}

In the following, we first review some related work in Section \ref{sec:related}, and then explain the details of our framework in Section \ref{sec:model}. We describe the three offline experimental results to verify the performance of the proposed approach in terms of rating prediction and explanation in Section \ref{sec:experiment}. In Section \ref{sec:discussion}, we will analyze the results and make discussions about what we learned from the experiments. Finally, we conclude this work and provide our visions of the research in Section \ref{sec:future}.

%%%%%%%%%%%%%%%%%%%%%%%%%%%%%%%%%%%%%%%%%%%%%%%%%%%
\section{Related Work}\label{sec:related}
Collaborative filtering (CF) \cite{schafer2007collaborative} has been an important approach to modern personalized recommendation systems. Early collaborative filtering methods adopted intuitive and explainable methods, such as user-based \cite{resnick1994grouplens} or item-based \cite{sarwar2001item} collaborative filtering, which makes recommendation based on similar users or similar items. Later approaches to CF more and more advanced to more accurate but less transparent latent factor approaches, beginning from various matrix factorization algorithms \cite{koren2009matrix,mnih2008probabilistic,srebro2005maximum,takacs2008investigation}, to more recent deep learning and neural modeling approaches \cite{zheng2017joint,zhang2017joint,zhang2017deep}. Though effective in ranking and rating prediction, the latent nature of these approaches makes it difficult to explain the recommendation results to users, which motivated the research on explainable recommendation \cite{zhang2018explainable}.

%key philosophy of these approaches is to learn user/item representations and similarity networks from data for better user-item matching, which makes it difficult to understand how the prediction function/network works to generate the output results, because the results are usually produced by latent similarity matching instead of an explicit reasoning procedure.

%Matrix Factorization (MF) \cite{koren2009matrix} has been one of the most useful CF-based algorithms. This method maps user and item vectors into a latent factor space. These latent vectors represent some latent features of users and items. The inner product of their latent vectors are treated as the rating prediction of a user to a specific item. However, this method cannot have a good performance when the matrix is sparse. 

Researchers have explored various approaches towards model-based explainable recommendation. Since user textual reviews are informative and better reflect user preferences, a lot of research explored the possibility of incorporating user reviews to improve the recommendation quality \cite{mcauley2013hidden,almahairi2015learning, zheng2017joint,ling2014ratings, xu2014collaborative} and recommendation explainability \cite{zhang2014explicit,ren2017social,wang2018explainable,chen2016learning,chen2018neural,li2017neural}, which helps to enhance the effectiveness, transparency, trustworthiness and persuasiveness of recommendation system \cite{herlocker2000explaining,zhang2014explicit}.

%They modeled textual reviews on word-level and map extracted feature words into latent topics to combine with latent factor model. Although these works show the improvement on rating accuracy by leveraging user textual reviews, the generated recommendation results are not explainable which is a well-known drawback of latent factor model \cite{koren2015advances}. To enhance the effectiveness, transparency, trustworthiness and persuasiveness of recommendation system, our work would not only make use of text to improve the rating performance but also generate free-text natural language explanations to enhance the recommendation quality.

%There are some research works make use of text information to generate text for recommendation system. 

Early approaches to explainable recommendation models generate explanations based on pre-defined explanation templates. For example, Zhang et al  \cite{zhang2014explicit} proposed an explicit factor model (EFM), which generates explanations by filling a sentence template with the item features that a user is interested in. 
%to do phrase-level sentiment analysis on user written textual reviews to extract user opinions. Then the model generates personalized recommendation results based on learned hidden features of a specific item. To make the recommended results explainable, the EMF would make use of predefined templates and fill in the extracted feature words to generate explanation sentences. 
However, generating explanation sentences in this way needs extensive human efforts to define various templates in different scenarios. Moreover, the predefined templates limit the expressive power of explanation sentences. Li et al \cite{li2017neural} leveraged neural rating regression and text generation to predict the user ratings and user-generated tips for recommendation, which helps to improve the prediction accuracy and the effectiveness of recommendation results. However, not all of the tips are of explanation purposes for the recommendations because they do not always explicitly comment about the product features.
%makes use of the user written tips information to improve recommendation performance. 
%Their model contains neural rating regression module and abstractive tips generation module. Both modules share the same user and item latent vector representation space. Although the rating accuracy is enhanced by leveraging tip information, the generated tip does not provide enough explanation to the recommendation results. It is because that one user generated tip only contains one sentence and it might be a general sentence without feature words of a specific item that the user cares about. 
To alleviate the problem, Costa et al \cite{costa2018automatic} attempted to train generation models based on user reviews and automatically generate fake reviews as explanations. One problem here is that not all of the sentences in the user reviews are appropriate for explanation purposes, because users may write sentences that are irrelevant to the corresponding item, which makes it difficult to generate explanations when the user reviews are too long with too much noise. Considering these deficiencies, we propose an auto-denoising mechanism for text generation and produce personalized natural language explanations for personalized recommendations.

Recently, deep neural network models have been used in various natural language processing tasks, such as question answering \cite{zhou2015answer} and text summarization \cite{nallapati2016abstractive}. A well trained neural network could learn lower-dimension dense representations to capture grammatical and semantical generalizations \cite{gatt2018survey}. This property of neural network is useful for natural language generation tasks. Recurrent neural network (RNN) \cite{mikolov2010recurrent} has shown notable success in sequential modeling tasks. The long short-term memory unit (LSTM) \cite{hochreiter1997long} and gated recurrent unit (GRU) \cite{cho2014properties} are among the most commonly used neural networks for natural language modeling to avoid the gradient vanishing problem when dealing with long sequences. A demonstration of potential utility of recurrent networks for natural language generation was provided by \cite{sutskever2011generating}, which used a character-level LSTM model for the generation of grammatical English sentences. Character-level models can obviate the Out-of-Vocabulary(OOV) problem and reduce the vector representation spaces for language modeling. However, they are generally outperformed by word-level models \cite{mikolov2012subword}. Considering the performance of these two modeling strategies, our proposed approach, in particular, works on word-level with GRU to generate natural language explanations.

%explainable important and what can be good explanation %\cite{Tintarev:2007:EER:1297231.1297259}

\section{The Framework}\label{sec:model}
\subsection{Overview}
An explainable recommender system can not only give an accurate prediction of rating score by given a user and an item, but also generate explanations to interpret the recommendation results. In our framework, we have two major modules: a rating prediction module and a natural language explanation generation module. Both modules take shared user and item latent factors as input. Since the input space is shared, the extra information can be utilized from the other module during training process to improve the general performance of our framework. During the testing stage, only user and item latent factors as well as the extracted feature words information are provided. 

At the training stage, the training data consists of users, items, user generated reviews and ratings. We use $\mathcal{X}$ to represent the training dataset; $\mathcal{U}$ and $\mathcal{I}$ are user set and item set respectively; $Review$ is the set of sentences in the user generated reviews; $\mathcal{R}$ represents the set of user ratings; $\mathcal{K}$ is the feature words set, where $\mathcal{K}$ is the subset of vocabulary $\mathcal{V}$. We have $\mathcal{X}=\{\mathcal{U}, \mathcal{I}, Review,\mathcal{R}, \mathcal{K}\}$. 
The key notations in this paper are listed in Table \ref{t:notation}. 

In the rating regression module, only the user latent factors $\textbf{U}$ and item latent factors $\textbf{V}$ are given as the input. Then the multi-layer perceptron would project these latent factors into a single value as the rating prediction. After that, we calculate the mean square error loss and optimize the loss function.

\begin{figure*}[t]
\centering
\includegraphics[scale=0.5]{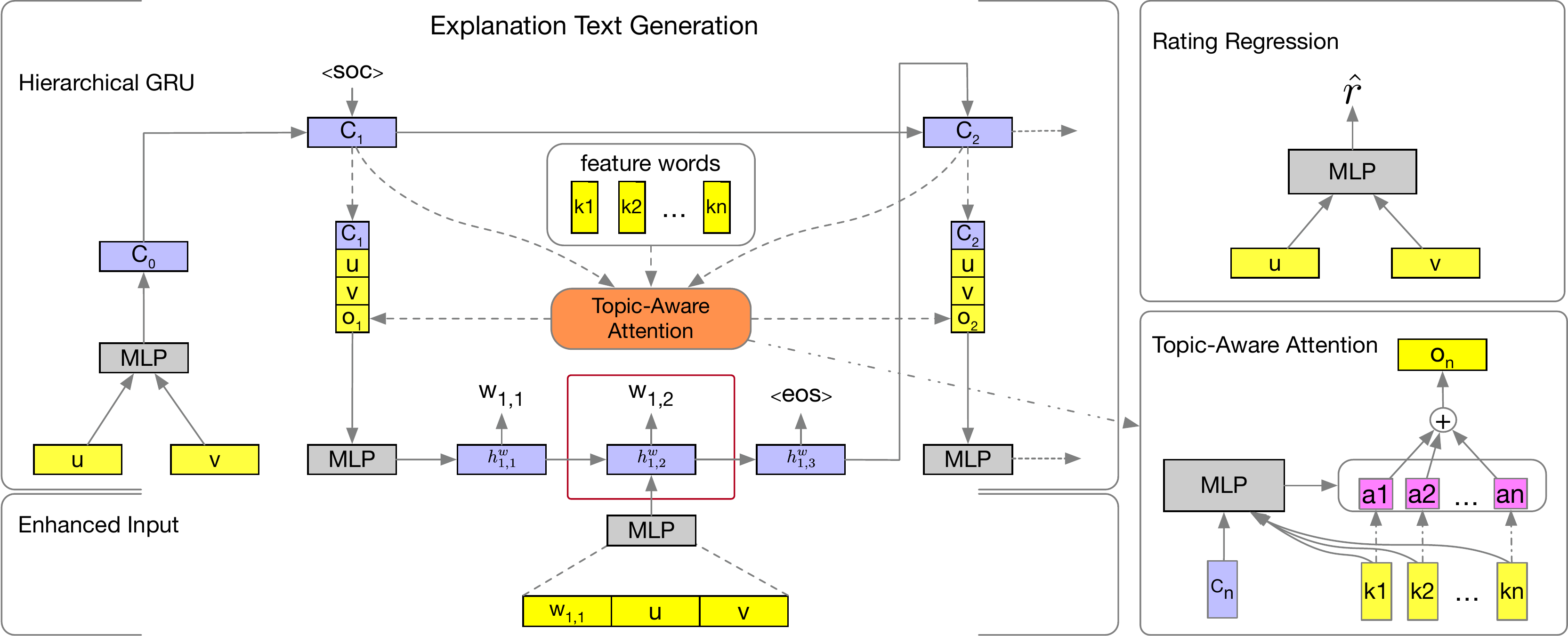}
\caption{Overview of our HSS model. There are two major modules---explanation text generation module and rating regression module. The yellow boxes represent latent factors, such as user latent factor u, item latent factor v, word vector w; blue boxes represent hidden states; gray boxes represent multi-layer perceptron; pink boxes represent the attention weights for each feature word vector.}
\label{fig:model}
\vspace{-15pt}
\end{figure*}

In the personalized natural language explanation generation module, we design a hierarchical GRU to map the user and item latent factors into a sequence of words. The overview of our framework is shown in Figure \ref{fig:model}. The hierarchical GRU contains a context GRU and a sentence GRU. Context GRU is used to generate the initial hidden state for sentence GRU to generate the sequence of words. The attention model is employed to improve the personalization of the generated sentences. It can be interpreted as which feature word or feature word should we pay more attention to when generating the current sentence. Since not all the words in the vocabulary are good for explanations and not all the feature words are suitable for each specific user item pair, we expect the model to learn to generate a more related and personalized explanation sentences by applying attention model. Moreover, we design an auto-denoising mechanism by applying a supervised factor on the corresponding generated sentence loss function. The key point here is that we believe the sentence with higher proportion of feature words would be more important for training the model. The effect of those sentences with less or zero proportion of feature words would automatically be weakened during training process by applying zero or very small supervised factor on their loss function. 

Finally, all the neural network parameters, user and item latent factors, word embedding in both modules are learned by a multi-task learning approach. The model can be trained through back-propagation algorithms.

\begin{table}[t]
\caption{A summary of key notations in this work.}\label{t:notation}
\begin{tabular}{|c|l|}
\hline
$\textbf{Notation}$ & \multicolumn{1}{c|}{$\textbf{Explanation}$} \\ \hline
$\mathcal{X}$ & \begin{tabular}[c]{@{}l@{}}training dataset\end{tabular} \\ \hline
$\mathcal{U}$ & user set \\ \hline
$\mathcal{I}$ & item set \\ \hline
$\mathcal{V}$ & \begin{tabular}[c]{@{}l@{}}vocabulary\end{tabular} \\ \hline
$\mathcal{K}$ & feature words set \\ \hline
$\mathcal{S}$ & The set of generated sequences \\ \hline
$Review$ & The set of sentences in the reviews \\ \hline
$\mathcal{R}$ & The set of user ratings \\ \hline
$\textbf{U}$ & The set of user latent factors\\ \hline
$\textbf{V}$ & The set of item latent factors \\ \hline
$\textbf{u}$ & user latent factor\\ \hline
$\textbf{v}$ & item latent factor\\ \hline
$\textbf{k}$ & feature word embedding\\ \hline
$\textbf{o}$ & attentive feature-aware vector\\ \hline
$\Theta$ & The set of neural network parameters \\ \hline
$\beta_i$ & The supervised factor of the i-th sentence \\ \hline
$d$ & \begin{tabular}[c]{@{}l@{}}latent factor dimension\end{tabular} \\ \hline
$r_{u,i}$ & rating of user $u$ to item $i$ \\ \hline
$tanh$ & hyperbolic tangent activation function \\ \hline
$\sigma$ & sigmoid activation function \\ \hline
$\phi$ & rectified linear unit activation function \\ \hline
$\varsigma$ & softmax function \\ \hline
\end{tabular}
\end{table}

\subsection{Neural Rating Regression}
The goal of doing neural rating regression is to make rating predictions by given user and item latent factors. We borrow the idea from paper \cite{li2017neural} which is to learn a function $f_r(\cdot)$ to project user latent factors $\textbf{U}$ and item latent factors $\textbf{V}$ to rating scores $\hat{\textbf{r}}$. Here $f_r(\cdot)$ is represented as a multi-layer perceptron (MLP):
\begin{equation}
    \hat{\textbf{r}} = \text{MLP}(\textbf{U},\textbf{V})
\end{equation}
where $\textbf{U}\in \mathbb{R}^{d\times m}$ and $\textbf{V}\in \mathbb{R}^{d\times n}$ are in different latent vector spaces; $m$ is the number of users and $n$ is the number of items; $d$ is the latent factor dimension for both user and item representations. We first map user and item latent factors into a hidden state:
\begin{equation}
    \textbf{h}^r = tanh(\textbf{W}_{uh}^r \textbf{u} + \textbf{W}_{vh}^r \textbf{v}+\textbf{b}_h^r)
\end{equation}
where $\textbf{W}_{uh}^r \in \mathbb{R}^{d\times d}$ and $\textbf{W}_{vh}^r \in \mathbb{R}^{d\times d}$; $\textbf{b}_h^r\in \mathbb{R}^{d\times 1}$ is the bias term. We add more layers and use tanh activation function to do non-linear transformation to improve the performance of rating prediction:
\begin{equation}
    \textbf{h}_l^r = tanh(\textbf{W}_{hh_l}^r \textbf{h}_{l-1}^r + \textbf{b}_{h_l}^r)
\end{equation}
where $\textbf{W}_{hh_l}^r\in \mathbb{R}^{d\times d}$ is a mapping matrix; $l$ is the index of hidden layer. We denote the last hidden layer as $\textbf{h}_L^r$. The output layer maps the last hidden state into a predicted rating score $\hat{r}$:
\begin{equation}
    \hat{r}=\textbf{W}_{hh_L}^r\textbf{h}_L^r+b_{h_L}^r
\end{equation}
where $\textbf{W}_{hh_L}^r\in \mathbb{R}^{1\times d}$. The objective function of this rating regression problem is defined as:
\begin{equation}
    \mathcal{L}^r = \frac{1}{|\mathcal{X}|}\sum_{u\in \mathcal{U}, i\in \mathcal{I}}(\hat{r}_{u,i}-r_{u,i})^2
\end{equation}
$\hat{r}_{u,i}$ is the predicted rating score by user $u$ given item $i$ and $r_{u,i}$ is the corresponding ground truth. We can optimize this objective function to learn neural network parameters $\Theta$ as well as user and item latent representations $\textbf{U}$ and $\textbf{V}$.

\subsection{Personalized Natural Language Explanation Generation}
The key point of doing this work is to generate personalized natural language explanations. Although some research works have already implemented deep neural models to generating reviews \cite{dong2017learning} or tips \cite{li2017neural}, not many researchers work on explanation generation. In this section, we will introduce: 1) auto-denosing strategy; 2) feature-aware attention for personalized explanation generation; 3) hierarchical GRU model for sentence generation.

\subsubsection{Auto-denoising}
User review usually contains multiple sentences. However, not all of them are good representations of user's purchase intention. Our goal is to promote the quality of generated explanation text by introducing a supervised factor to control the training process, so that our model can learn from those more important sentences while ignoring those useless sentences. To implement this idea, we first extract all the feature words by using toolkit Sentires\footnote{\url{https://github.com/evison/Sentires}}\cite{zhang2014explicit,zhang2014users}, represented as $\mathcal{K}$ and $\mathcal{K}\subseteq \mathcal{V}$, from the data set. Then the supervised factor of the $i$-th sentence in the review is calculated as:
\begin{equation}
    \beta_i = \frac{N_k^i}{N_w^i}
\end{equation}
where $N_k^i$ is the number of feature words in the $i$-th sentence; $N_w^i$ is the total number of words in the $i$-th sentence. We can multiply this supervised factor to the loss function of this sentence to control the training process. We believe that the sentence with higher feature words proportion would be more important. The effect of those sentences with lower or zero proportion of feature words would be automatically weakened by multiplying zero or quite small factor on their loss function. 

\subsubsection{Feature-aware Attention}
\label{attention}
Feature words are the words which describe the features of a product. For example, "memory", "screen", "sensitivity" can be feature words in electronics dataset. However, "use", "good", "day" are not feature words, since they are not used to describe the feature of an item. Since users may pay different attention to these feature words and each product may only relate to some feature words, inspired by \cite{xing2017topic}, we implement a feature-aware attention mechanism to improve the personality. Mathematically, given a hidden state $\textbf{h}_t$ and the $i-$th feature word embedding $\textbf{k}_i$, the attention score of the feature word $\textbf{k}_i$ at time $t$ is computed as:
\begin{equation}
\begin{split}
    \textbf{x}_i &= \textbf{h}_t;\textbf{k}_i \\
    a(i,\textbf{h}_t) &= \textbf{w}_2^T \phi(\textbf{W}_1^a\textbf{x}_i + \textbf{b}_1^a)+b_2^a
\end{split}
\end{equation}
where $\textbf{x}_i\in \mathbb{R}^{2d\times 1}$ is the concatenation of the hidden vector at time $t$ and the $i$-th feature word vector; $\textbf{W}_1^a\in \mathbb{R}^{d\times 2d}$ is the mapping matrix for the first layer network; $\textbf{b}_1^a\in \mathbb{R}^{d\times 1}$ is the first layer bias; $\textbf{w}_2\in \mathbb{R}^{d\times 1}$ and $b_2^a\in \mathbb{R}$ are the neural parameters for the second layer; $\phi(\cdot)$ is the ReLU activation function which is defined as:
\begin{equation}
\phi(x) = max(0, x)
\end{equation}
The final attention weights are obtained by normalizing above attentive scores using softmax, which can be interpreted as how much attention do we pay to the feature word in term of corresponding hidden state during the training process. %(need to be modified)
\begin{equation}
\alpha(i,\textbf{h}_t) = \frac{exp(a(i,\textbf{h}_t))}{\sum_{i=1}^{|\mathcal{K}|}exp(a(i, \textbf{h}_t))}
\end{equation}
Finally, the attentive feature-aware vector at time $t$ is calculated as:
\begin{equation}
    \textbf{o}_t = \sum_{i=1}^{|\mathcal{K}|}\alpha(i, \textbf{h}_t)\textbf{k}_i
\end{equation}

This attentive feature-aware vector will be used to compute the initial hidden state for generating the $i$-th sentence in $GRU_{wrd}$, which will be introduced in the following subsection. 

\subsubsection{Context-level GRU ($GRU_{ctx}$)}
As shown in Figure \ref{fig:model}, the review sentences are generated by $GRU_{wrd}$, which will be introduced in the next subsection. However, the initial hidden states are given by $GRU_{ctx}$. By leveraging this hierarchical recurrent neural network, we can generate multiple sentences by given one pair of user and item latent factors. Since each generated sentence has its own loss function, we are able to apply the auto-denoising strategy mentioned above to reduce the effects of unrelated sentences in the user generated reviews during training process. 

Suppose that for each user and item pair, there are $n$ sentences in the review. Then we have $n$ context representations. 
\[C = \{\textbf{C}_1,\textbf{C}_2,\ldots,\textbf{C}_n\}\]
We use $C$ to denote the collection of all the context representations and $\textbf{C}_i$ denotes a specific context representation. When a sentence is generated, the context representation would be updated by the following equation:
\begin{equation}
\textbf{C}_n = GRU_{ctx}(\textbf{C}_{n-1}, \textbf{h}_{n-1,L}^w)
\end{equation}
$\textbf{C}_{n-1}$ is the previous context representation; $\textbf{h}_{n-1,L}^w$ is the last hidden state of the $GRU_{wrd}$. Then the $GRU_{ctx}$ state is updated by the following operations:
\begin{equation}
\begin{split}
\textbf{r}_n^c &= \sigma(\textbf{W}_{hr}^c\textbf{h}_{n-1,L}^w+\textbf{W}_{cr}^c\textbf{C}_{n-1}+\textbf{b}_r^c) \\
\textbf{z}_n^c &= \sigma(\textbf{W}_{hz}^c\textbf{h}_{n-1,L}^w+\textbf{W}_{cz}^c\textbf{C}_{n-1}+\textbf{b}_z^c) \\
\textbf{g}_n^c &= \tanh(\textbf{W}_{hg}^c\textbf{h}_{n-1,L}^w+\textbf{W}_{cg}^c(\textbf{r}_{n}^c\odot\textbf{C}_{n-1})+\textbf{b}_g^c) \\
\textbf{C}_n &= \textbf{z}_n^c\odot\textbf{C}_{n-1}+(1-\textbf{z}_n^c)\odot\textbf{g}_n^c
\end{split}
\end{equation}

To start the whole process, we utilize the user latent factor $\textbf{u}$ and item latent factor $\textbf{v}$ to initialize the first hidden state $\textbf{C}_0$. 
\begin{equation}
\textbf{C}_0 = \phi(\textbf{W}_u^{c_0}\textbf{u}+\textbf{W}_v^{c_0}\textbf{v}+\textbf{b}^{c_0})
\end{equation}

\subsubsection{Word-level GRU ($GRU_{wrd}$)}
This part is to generate the words for explanation sentences. The main idea can be descripbed as follow:
\begin{equation}
p(w_{n,t} | w_{n,1},w_{n,2},\ldots,w_{n,t-1})=\varsigma(\textbf{h}_{n,t}^w)
\end{equation}
where $w_{n,t}$ is the $t$-th word of the $n$-th review sentence. $\varsigma(\cdot)$ is the softmax function which is defined as follow:
\begin{equation}
\label{eq:softmax}
\varsigma(x_i) = \frac{e^{x_i}}{\sum_je^{x_j}}
\end{equation}
$\textbf{h}_{n,t}^w$ is the sequence hidden state of the $n$-th sentence at the time $t$. It depends on the previous hidden state $\textbf{h}_{n,t-1}^w$ and the current input $\textbf{w}_{n,t}$:
\begin{equation}
\textbf{h}_{n,t}^w = f(\textbf{h}_{n,t-1}^w, \textbf{w}_{n,t})
\end{equation}
The $f(\cdot)$ can be LSTM, GRU or Vanilla RNN. Here we utilize GRU for efficiency consideration. The states are updated by following operations:
\begin{equation}
\begin{split}
\textbf{r}_{n,t}^w &= \sigma(\textbf{W}_{wr}^w\textbf{w}_{n,t}+\textbf{W}_{hr}^w\textbf{h}_{n,t-1}^w+\textbf{b}_r^w) \\
\textbf{z}_{n,t}^w &= \sigma(\textbf{W}_{wz}^w\textbf{w}_{n,t}+\textbf{W}_{hz}^w\textbf{h}_{n,t-1}^w+\textbf{b}_z^w) \\
\textbf{g}_{n,t}^w &= \tanh(\textbf{W}_{wg}^w\textbf{w}_{n,t}+\textbf{W}_{hg}^w(\textbf{r}_{n,t}^w\odot\textbf{h}_{n,t-1}^w)+\textbf{b}_g^w) \\
\textbf{h}_{n,t}^w &= \textbf{z}_{n,t}^w\odot\textbf{h}_{n,t-1}^w+(1-\textbf{z}_{n,t}^w)\odot\textbf{g}_{n,t}^w
\end{split}
\end{equation}
where $\textbf{r}_{n,t}^w$ is the reset gate; $\textbf{z}_{n,t}^w$ is the update gate; $\odot$ represents element-wise multiplication; $\textit{tanh}$ denotes hyperbolic tangent activation function; $\textbf{w}_{n,t}$ can simply to be the vector representation of the word $w_{n,t}$, which is the word in the $n$-th sentence in the review at time $t$. However, we expect to bring more personalized information into the text generation model. Inspired by \cite{tang2016context} we concatenate word embedding of word $w$ at time $t$ with user embedding $\textbf{u}$ and item embedding $\textbf{v}$, to get an enhanced input embedding $\textbf{s}_{n,t}$. Then we feed this embedding into a multi-layer perceptron to produce input vector $\textbf{w}_{n,t}$:
\begin{equation}
\begin{split}
    \textbf{s}_{n,t} &= \textbf{e}_{n,t};\textbf{u};\textbf{v} \\
    \textbf{h}_s &= \phi(\textbf{W}_s\textbf{s}_{n,t}+\textbf{b}_s) \\
\end{split}
\end{equation}
where $\textbf{e}_{n,t}$ is the vector representation of word $w$ in the $n$-th sentence at time $t$; $\textbf{h}_s$ is the hidden state after doing non-linear transformation on the enhanced embedding. We can add more layers and finally feed the output of the last layer hidden state $\textbf{h}_L^s$ into an output layer to get the input vector $\textbf{w}_{n,t}$:
\begin{equation}
    \textbf{w}_{n,t} = \textbf{W}_L^s\textbf{h}_L^s + \textbf{b}_L^s
\end{equation}
where $\textbf{W}_s \in \mathbb{R}^{d\times 3d}$, $\textbf{W}_L^s\in \mathbb{R}^{d\times d}$; $\textbf{b}_s$ and $\textbf{b}_L^s$ are in $\mathbb{R}^d$.

To start the explanation sentences generation process, we need an initial hidden state. We use the output of $GRU_{ctx}$ $\textbf{C}_n$, user latent factor $\textbf{u}$, item latent factor $\textbf{v}$ and the $i$-th sentence feature-aware attentive context vector $\textbf{o}_n$ together to compute the initial hidden state $\textbf{h}_{n,0}^w$:
\begin{equation}
\label{eq:hn0}
\textbf{h}_{n,0}^w = {\textbf{W}_{n,2}^{i}}^T\phi(\textbf{W}_{n,1}^{i}(\textbf{C}_n; \textbf{u}; \textbf{v}; \textbf{o}_n)+\textbf{b}_{n,1}^i)+\textbf{b}_{n,2}^i
\end{equation}
where $\textbf{W}_{n,1}^i\in \mathbb{R}^{d\times 4d}$, $\textbf{b}_{n,1}^i\in \mathbb{R}^{d\times 1}$, $\textbf{W}_{n,2}^i\in \mathbb{R}^{d\times d}$, $\textbf{b}_{n,2}^i\in \mathbb{R}^{d\times 1}$. The feature-aware attentive context vector $\textbf{o}_n$ is calculated as described in subsection \ref{attention}, where the hidden state $\textbf{h}_t$ is replaced with $\textbf{C}_n$ and the feature-aware attentive context vector is represented as $\textbf{o}_n$ instead of $\textbf{o}_t$. This can be interpreted as how much attention do the model pay to the feature words when generating the $n$-th explanation sentence. The equation (\ref{eq:hn0}) uses two layers neural network to calculate initial hidden state for $GRU_{wrd}$. You can choose to add more layers here.

By obtaining the $\textbf{h}_{n,0}^w$, GRU can conduct the sequence decoding process. After obtaining all the hidden states of the sequences, we then feed them into a final output layer to predict the word sequence in the review. 
\begin{equation}
\hat{\textbf{y}}_{t+1} = \varsigma(\textbf{W}_{h}^w\textbf{h}_t^w + \textbf{b}^w)
\end{equation}
$\varsigma(\cdot)$ is softmax function which was defined in Equation \ref{eq:softmax}; $\textbf{h}_t^w \in \mathbb{R}^{d\times l}$ is the hidden state matrix, where $l$ is the length of the sequence; $\textbf{W}_h^w \in \mathbb{R}^{|\mathcal{V}|\times d}$; $\hat{\textbf{y}}_{t+1}$ can be considered as a multinomial distribution over vocabulary $\mathcal{V}$ on review text. Then the model can generate the next word $w_{t+1}^*$ from $\hat{\textbf{y}}_{t+1}$ by selecting the one with the largest probability. Here we use $w_i$ to indicate the $i$-th word in vocabulary. Then we have 
\begin{equation}
w_{t+1}^* = \mathop{\mathrm{argmax}}_{w_i \in V} \hat{\textbf{y}}_{t+1}^{(w_i)}
\end{equation}

To train the model, we use Negative Log-Likelihood as the loss function. Our goal is to make the words in the review have higher probabilities than others. Here $I_w$ is the index of word $w$ in the vocabulary $\mathcal{V}$. The loss function of the $i-$th sentence is represented as:
\begin{equation}
\mathcal{L}_i^s = - \sum_{w\in Review}\log \hat{\textbf{y}}^{(I_w)}
\end{equation}

In the testing stage, we introduce beam search to search for the best sequence $s^*$ with maximum log-likelihood. \begin{equation}
    s^* = \mathop{\mathrm{argmax}}_{s\in\mathcal{S}}\sum_{w\in s}\log \hat{\textbf{y}}^{(I_w)}
\end{equation} 
$\mathcal{S}$ is the set of generated sequences. $|\mathcal{S}|$ is the beam size.

\subsection{Multi-task Learning}
The framework contains two major modules. We integrate both parts into one multi-task learning process. The final objective function is defined as:
%WHY %
\begin{equation}
\mathcal{J} = \min_{\textbf{U},\textbf{V},\textbf{E},\Theta}\big(\mathcal{L}^r + \sum_{i=1}^{|Review|}\beta_i\mathcal{L}_i^s + \lambda(||\textbf{U}||_2^2 + ||\textbf{V}||_2^2 + ||\Theta||_2^2)\big)
\end{equation}
where $\textbf{E}\in \mathbb{R}^{d\times |\mathcal{V}|}$ is the word embedding matrix; $\Theta$ is the set of neural parameters; $\lambda$ is the penalty weight; $\mathcal{L}^r$ is the rating regression loss function; $\beta_i$ is the supervised factor of the $i$-th sentence; $\beta_i\mathcal{L}_i^s$ is the weighted loss function of the $i$-th generated sentence for auto-denoising.

\section{Experiments}\label{sec:experiment}
\subsection{Datasets}
Our datasets are built upon Amazon 5-core~\footnote{\url{http://jmcauley.ucsd.edu/data/amazon}}~\cite{mcauley2015image} which includes user generated reviews and metadata spanning from May 1996 to July 2014 without duplicated records. The dataset covers 24 different categories and we select \textbf{Electronics} and \textbf{Beauty} two datasets to cover different domains and different scales in our experiment. Instead of using original 5-core version, we filter the dataset by selecting the users who has at least 10 shopping records. The reason of doing this filtering operation is that the model would not be well trained to learn the personalized preference for those users with very few reviews. After the original 5-core data is filtered, we move those records with less item frequencies into training set to get avoid of cold start issue in testing stage. For review text pre-processing, we keep all the punctuation and numbers in the raw text and we do not remove long sentences by setting a length threshold. In other words, our dataset is noisy which is challenging for text generation models. The "Electronics" dataset contains 45,224 users, 61,687 items, 744,453 reviews and 434 extracted feature words; the "Beauty" dataset is a smaller dataset which contains 5,122 users, 11,616 items, 90,247 reviews and 518 extracted feature words. The statistical details of our datasets are in Table~\ref{t:dataset}.

We filter out the words with frequency lower than ten to build the vocabulary $\mathcal{V}$. Then the whole dataset is splitted into three subsets: training, validation and testing (80\%/10\%/10\%). 

\begin{table}[t]
\caption{Statistics of the datasets in our experiments.}\label{t:dataset}
\begin{tabular}{cccc}
\toprule
 &\textbf{Electronics} & \textbf{Beauty}\\ 
\midrule
\#Users &45,224 &5,122\\ 
\#Items &61,687 &11,616\\ 
\#Reviews &744,453 &90,247\\
\#features &434&518\\
$\mid\mathcal{V}\mid$ &20,568 &7,152\\
sparsity &99.999\% & 99.998\%\\
\bottomrule
\end{tabular}
\vspace{-20pt}
\end{table}

\subsection{Rating Regression Evaluation}
\subsubsection{\textbf{Baselines}}
To evaluate the performance of rating prediction, we compare our HSS model with three methods, namely BiasedMF, SVD++ and DeepCoNN. The first two methods only utilize the ratings information and the third method involves user generated review for rating prediction. 
\begin{itemize}
    \item \textbf{BiasedMF}~\cite{koren2009matrix}: Biased Matrix Factorization. It only uses rating matrix to learn two low-rank user and item matrices to do rating prediction. By adding biases into plain matrix factorization model, it is able to depict the independent interaction of a user or an item on a rating value.
    \item \textbf{SVD++}~\cite{koren2008factorization}: It extends Singular Value Decomposition by integrating implicit feedback into latent factor modeling.
    \item \textbf{DeepCoNN}\cite{zheng2017joint}: Deep Cooperative Neural Networks. This is a state-of-art deep learning method that exploits user reviews information to jointly model user and item. The author has shown that their model significantly outperforms some strong topic modeling based methods such as \textbf{HFT}\cite{mcauley2013hidden} and \textbf{CTR}\cite{wang2011collaborative}. We use the implementation by \cite{chen2018neural} in our experiments. 
\end{itemize}
\subsubsection{\textbf{Evaluation Metric}}
To evaluate the performance of rating prediction, we employ the well-known Root Mean Square Error (RMSE) as our evaluation metric. 
%Given a rating prediction $\hat{r}_{u,i}$ and the ground truth $r_{u,i}$, the RMSE is calculated as:
%\begin{equation}
%    RMSE = \sqrt{\frac{1}{N}\sum_{u\in\mathcal{U},i\in\mathcal{V}}(r_{u,i}-\Hat{r}_{u,i})^2}
%\end{equation}
%where $N$ is the total number of observations.

\subsection{\mbox{Explanation Sentence Generation Evaluation}}
\subsubsection{\textbf{Baseline}}
To evaluate the performance of text generation module, we compare our work with \textbf{Att2SeqA}\cite{li2017learning}. This work is to automatically generate product reviews by given user, item and corresponding rating information. Their model treat user, item and rating as attributes and encode the three attributes into latent factors through multi-layer perceptron. Then the decoder would take the encoded latent factor as the initial hidden state of LSTM for review generation. In our implementation, we also use the two-level stacked LSTM for text generation as the paper proposed. There are three reasons for choosing this model as our baseline:
\begin{itemize}
    \item \textit{Similar input}: both our \textbf{HSS} and their \textbf{Att2SeqA} would learn user and item latent factor as the input for text generation. The difference is that their model would take rating information as the direct input while our model would learn to predict the rating score.
    \item \textit{Use attention mechanism}: their model introduces attention mechanism to enhance the text generation quality while our model also uses attention model to improve the personality of the generated explanations. 
    \item \textit{Use review data}: both methods use user generated reviews for training the model. The difference is that their model is to learn from user written review to automatically generate fake reviews while our model is to generate explanation sentences. 
\end{itemize}
Considering these three reasons, we believe that this model is the most suitable and competitive model for comparison. 

There is another related model called \textbf{NRT} proposed in \cite{li2017neural}. In that paper, they also do rating regression and text generation simultaneously. However, their goal is to generate tips. The data source used by their model is the summary in Amazon dataset. The summary can be treated as the title of a user review. It only contains one short sentence expresses the general feeling of a user to a product such as "So good", "Excellent", "I don't like it". Since the summaries or tips are too general to depict the features of an item that a user is preferred, we cannot use summaries for training an explanation generation model. The \textbf{NRT} model is very useful for simulating user feelings on a specific item. However, considering the differences from data source and the designing purposes, we would not use this model as our baseline.

\subsubsection{\textbf{Evaluation Metrics}}
We use three evaluation metrics to evaluate generated explanation sentence quality: BLEU\cite{papineni2002bleu}, ROUGE\cite{lin2004rouge} and feature words coverage. 
\begin{itemize}
    \item \textit{BLEU}: this is a precision-based measure which is used for automatically evaluating machine generated text quality. It measures how well a machine generated text (candidate) matches a set of human reference texts by counting the percentage of n-grams in the machine generated text overlapping with the human references. The precision score for n-gram is calculated as:
    \[
        p_n=\frac{\sum_{C\in\{Candidates\}}\sum_{ngram\in C}Count_{clip}(ngram)}{\sum_{C'\in\{Candidates\}}\sum_{ngram'\in C'}Count(ngram')}
    \]
    where $Count_{clip}$ means that the count of each word in the machine generated text is truncated to not exceed the largest count observed in any single reference for that word. For more details, please refer to the paper \cite{papineni2002bleu}.
    \item \textit{ROUGE}: this is another classical evaluation metric for evaluating machine generated text quality. It is a recall-related measure which shows how much the words in the human reference texts appear in the machine generated text. The ROUGE-N is computed as:
    \[\text{ROUGE-N}=\frac{\sum_{S\in \{References\}}\sum_{ngram\in S}Count_{match}(ngram)}{\sum_{S\in \{References\}}\sum_{ngram\in S}Count(ngram)}\]
    where $Count_{match}(ngram)$ is the maximum number of n-grams co-occurring in a machine generated text and a set of human reference texts. In our experiments, we use recall, precision and F-measure of ROUGE-1(uni-gram), ROUGE-2(bi-gram), ROUGE-L(longest common subsequence) and ROUGE-SU4(skip gram) to evaluate the quality of generated explanation sentences. We use the standard option~\footnote{ROUGE-1.5.5.pl -n 4 -w 1.2 -m -2 4 -u -c 95 -r 1000 -f A -p 0.5} for evalutaion.
    \item \textit{Feature words coverage}: this measure is to reflect how well our model can capture the user personalized preferences. Assuming that the number of feature words in the human reference texts is $N_r$ and the number of covered feature words in the machine generated sentences is $N_c$, the feature words coverage is calculated as:
    \[Coverage_{feature} = \frac{N_r}{N_c}\]
    We use this measure to reflect how well our model generated sentences can capture the users personalized preferences. In the meanwhile, this is also the measure we use to evaluate the explainability of the generated explanation sentences.
\end{itemize}
\subsection{Experimental Settings}
In our HSS model, we use 300 as the dimension of user and item latent factors. The dimension of hidden size and word vector are set to 300. The number of layers for rating regression model is 4 and for explanation generation is 3. The training batch size is 100. We add gradient clip on $GRU_{ctx}$ and $GRU_{sen}$ by setting the norm of gradient clip to 1.0. The L2 regularization weight parameter $\lambda=0.001$, dropout rate is 0.1. The beam size is set to 4 for both our model and the baseline model. All the linear layers parameter matrices are initialized from a normal distribution with mean is 0, standard deviation is 0.05. The parameter matrices in GRU are initialized with random orthogonal matrices. We set the learning rate to 0.002. The opitmizer is SGD with momentum equal to 0.9. 

For the Att2SeqA model, we set the user, item and rating latent factor to 64. The hidden size and word vector size is 512. Training batch size is set to 100. The dropout rate is 0.2. Learning rate is 0.002. The optimizer is RMSprop with alpha equal to 0.95.

For both HSS and Att2SeqA, we set the length of the generated sequence to 100 but only keep the first two sentences by searching for the tag of the end of sentence "EOS". The remainder of the generated sequence would be discarded. We use these two sentences for evaluation. The reason why we do this is because a shorter explanation would be easier for users to get the point of a specific item quickly. However, if the explanation is too short, for example one sentence, that one sentence may not cover enough information to improve the recommendation quality. We think two is a reasonable good length for explanation sentences. 

\section{Results and Discussions}\label{sec:discussion}
\subsection{Ratings}
Our HSS model can not only generate natural language explanation sentences, but also provide predicted rating scores. The results of rating prediction of our model and baseline models are given in Table \ref{t:rmse}. It shows that our model can outperforms all the baselines on Beauty dataset. On Electronics dataset, the RMSE score of our HSS is better than BiasedMF and SVD++. Although the performance is not better than the state-of-art model DeepCoNN, the result is still comparable. In general, the topic-based deep neural network model DeepCoNN and HSS are better than tradition collaborative filtering based methods. It is because that DeepCoNN and HSS takes user reviews to improve the representation ability of user and latent factors, while the traditional methods only use rating information. 

The difference between our HSS and DeepCoNN is the way of using the reivew data. In our HSS, we use GRU to learn to generate a sequence of words. The review data is used for maximizing the log likelihood of generated words. The DeepCoNN maps the user review content into a set of word embedding. Then pass the word embedding into convolution layers, max-pooling layer and fully connected layers to map the word embedding into a rating score. Although the way of using the review data is different, the experimental results on both models show that it is helpful to make use of user review information to improve the recommendation performance. 

\begin{table}[t]
\caption{RMSE values for rating regression}\label{t:rmse}
\vspace{-10pt}
\begin{tabular}{ccc}
\toprule
 &\textbf{Electronics} &\textbf{Beauty}\\ 
\midrule
BiasedMF &1.096 &1.030 \\ 
SVD++ &1.104 &1.034 \\ 
DeepCoNN &\textbf{1.089} &1.028\\
\textbf{HSS} &1.090 &\textbf{1.027}\\
\bottomrule
\end{tabular}
\vspace{-10pt}
\end{table}

\subsection{Personalized Explanation Sentence Generation Quality}

In order to evaluate the quality of generated sentences, we report recall, precision and F-measure of ROUGE-1, ROUGE-2, ROUGE-L, ROUGE-SU4. The results are shown in Table \ref{t:rouge_elec}--\ref{t:rouge_beauty}. According to the results, our model almost outperforms the baseline model on all the measures and only the recall on ROUGE-SU4 is slightly lower than the baseline model. From the results we can see that both models achieve good recall scores on all the measures except for ROUGE-2. One possible reason is that both models employ attention mechanism during the sequence generation process. The experiment result reflect that by adding attention context vector on word generation process can help to generate the sentences which are more related to the user and the product. 

One difference between our model and the Att2SeqA model is that we implement the attention model by leveraging feature words. We believe that not all the feature words are related to a specific item and each user has their own preferred features. Considering this property in the e-commerce scenario, we calculate the attention weights on each of the feature word embedding with the context level hidden state on current time stamp. Then we apply the attention weights on each of the feature words embedding and integrate all the weighted embedding into a attentive context vector. This context vector represents how much attention do the model pay to each feature word when generating current sentence. However, Att2SeqA model obtains the attentive context vector with user, item and rating latent factors, which are the attributes as mentioned in the paper~\cite{li2017learning}. Then the author combines this context vector with the output of GRU on each time stamp to predict the next word. Since their attention mechanism is not for improving the feature word coverage, our model get much higher score on feature words coverage as shown in Table \ref{t:bleu}. In another word, to do attention on feature words do help the model to cover more feature words in the generated sentences.

Another observation is that our model gives much higher precision score than the baseline model. It means that our model generated sequences can hit much more words in human reference texts than those generated by Att2SeqA. As shown in Table \ref{t:bleu}, the BLEU-score, which is a precision-based metric for text generation evaluation, also gives a higher score to HSS than Att2SeqA. 

\begin{table}[t]
\caption{BLEU-1 (B-1), BLEU-4 (B-4) and Feature words coverage (FC) on Electronics and Beauty dataset (in percentage)}\label{t:bleu}
\vspace{-10pt}
\centering\begin{tabular}{|c|c|c|c|c|c|c|}
\hline
\multirow{2}{*}{} &\multicolumn{3}{|l|}{\textbf{Electronics}} & \multicolumn{3}{|l|}{\textbf{Beauty}} \\ 
\cline{2-7}
&B-1 &B-4 &FC &B-1 &B-4 &FC \\
\hline
 Att2SeqA &7.32 &2.17 &2.16 &8.54 &1.61 &1.69\\ 
 \hline
 \textbf{HSS} &\textbf{12.36} &\textbf{4.17} &\textbf{6.74} &\textbf{9.55} &\textbf{3.49} &\textbf{6.05}\\
 \hline
\end{tabular}
\vspace{-10pt}
\end{table}

\begin{table*}[t]
\caption{ROUGE score on Electronics dataset (in percentage)}\label{t:rouge_elec}
\vspace{-10pt}
\centering\begin{tabular}{|c|c|c|c|c|c|c|c|c|c|c|c|c|}
\hline
\multirow{2}{*}{} &\multicolumn{3}{|l|}{\textbf{ROUGE-1}} & \multicolumn{3}{|l|}{\textbf{ROUGE-2}} &\multicolumn{3}{|l|}{\textbf{ROUGE-L}} & \multicolumn{3}{|l|}{\textbf{ROUGE-SU4}} \\ 
\cline{2-13}
&recall &precision &F1 &recall &precision &F1 &recall &precision &F1 &recall &precision &F1 \\
\hline
 Att2SeqA &22.80 &7.79 &10.19 &0.45 &0.14 &0.18 &19.93 &6.77 &8.85 &9.26 &1.07 &1.38 \\ 
 \hline
 \textbf{HSS} &\textbf{26.76} &\textbf{15.72} &\textbf{18.36} &\textbf{3.01} &\textbf{1.77} &\textbf{2.05} &\textbf{22.51} &\textbf{13.31} &\textbf{15.47} &\textbf{9.69} &\textbf{3.51} &\textbf{4.10}\\ 
 \hline
\end{tabular}
%\vspace{-5pt}
\end{table*}

\begin{table*}[t]
\caption{ROUGE score on Beauty dataset (in percentage)}\label{t:rouge_beauty}
\vspace{-10pt}
\centering\begin{tabular}{|c|c|c|c|c|c|c|c|c|c|c|c|c|}
\hline
\multirow{2}{*}{} &\multicolumn{3}{|l|}{\textbf{ROUGE-1}} & \multicolumn{3}{|l|}{\textbf{ROUGE-2}} &\multicolumn{3}{|l|}{\textbf{ROUGE-L}} & \multicolumn{3}{|l|}{\textbf{ROUGE-SU4}} \\ 
\cline{2-13}
&recall &precision &F1 &recall &precision &F1 &recall &precision &F1 &recall &precision &F1 \\
\hline
 Att2SeqA &26.55 &8.67 &12.03 &0.70 &0.19 &0.27 &22.96 &7.57 &10.46 &\textbf{11.54} &1.31 &1.91 \\ 
 \hline
 HSS &\textbf{28.40} &\textbf{13.49} &\textbf{16.85} &\textbf{4.07} &\textbf{1.85} &\textbf{2.31} &\textbf{24.64} &\textbf{11.66} &\textbf{14.57} &11.43 &\textbf{2.73} &\textbf{3.48}\\
 \hline
\end{tabular}
\end{table*}

\begin{table*}[t]
\caption{Example of generated sentences. The feature words are marked in bold.}\label{t:sentence}
\vspace{-10pt}
\centering\begin{tabular}{c|c}
\hline
 Description & Explanation Sentences\\
\hline
 \textit{good explanation on Beauty}& The \textbf{bottle} is very \textbf{light} and the \textbf{smell} is very strong. \\
 \hline
\textit{good explanation on Electronics}& The \textbf{price} is great. The \textbf{sound quality} is great\\
\hline
 \textit{cover feature words but not fluent}& The \textbf{scent} is a good \textbf{product}. I have to use this \textbf{product}. I have used to use the \textbf{hair}.\\ 
 \hline
\textit{fluent but wrong description} & the \textbf{price} is a great. The \textbf{sound} is great\\
\hline
\end{tabular}
\vspace{-10pt}
\end{table*}

\subsection{Multi-sentence Generation Performance}
\begin{figure*}[h]
\centering     %%% not \center
\caption{ROUGE scores change on the number of generated sentences}
\vspace{-15pt}
\subfigure{\label{fig:rouge_recall}\includegraphics[width=.32\linewidth]{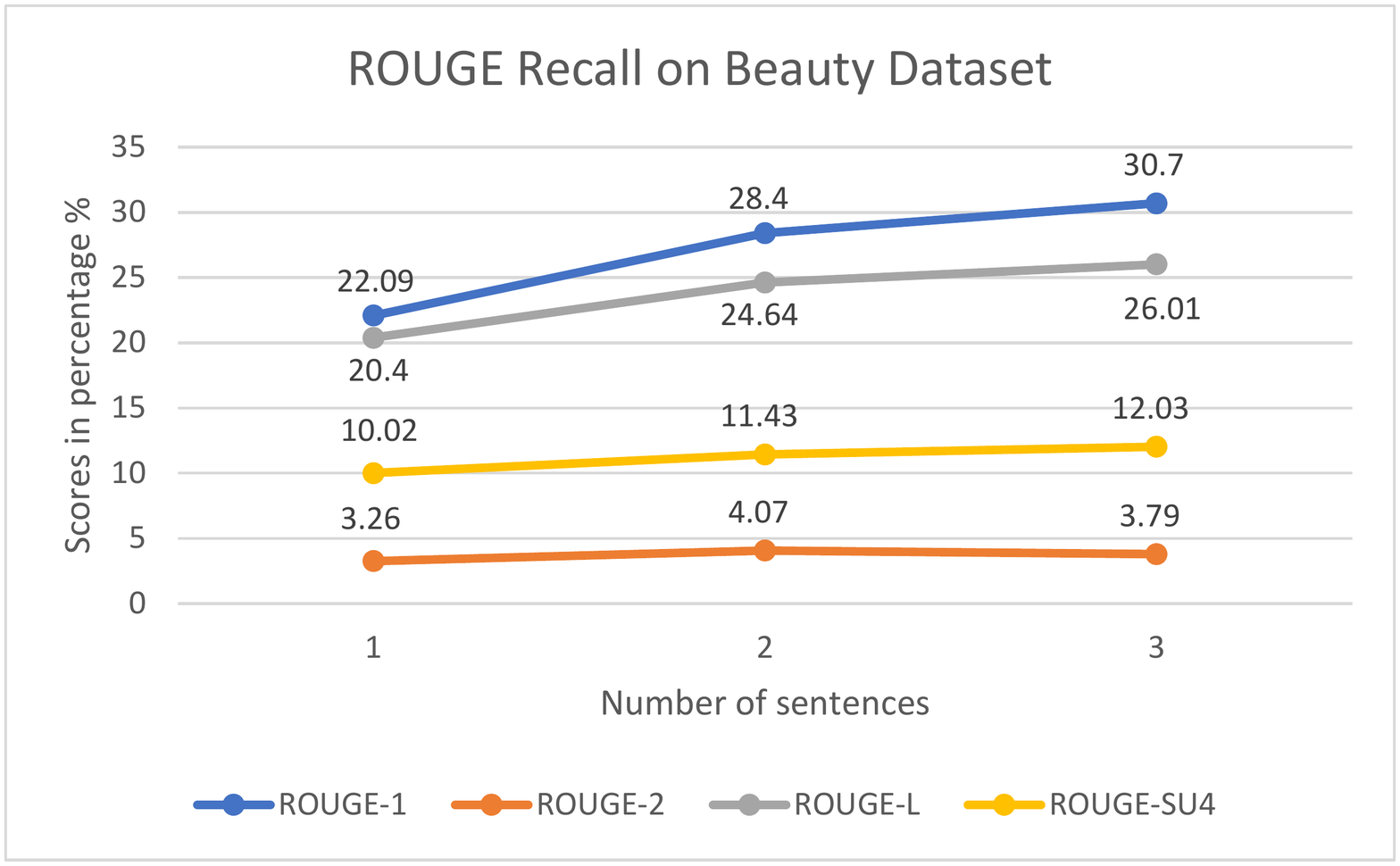}}
\subfigure{\label{fig:rouge_precision}\includegraphics[width=.32\linewidth]{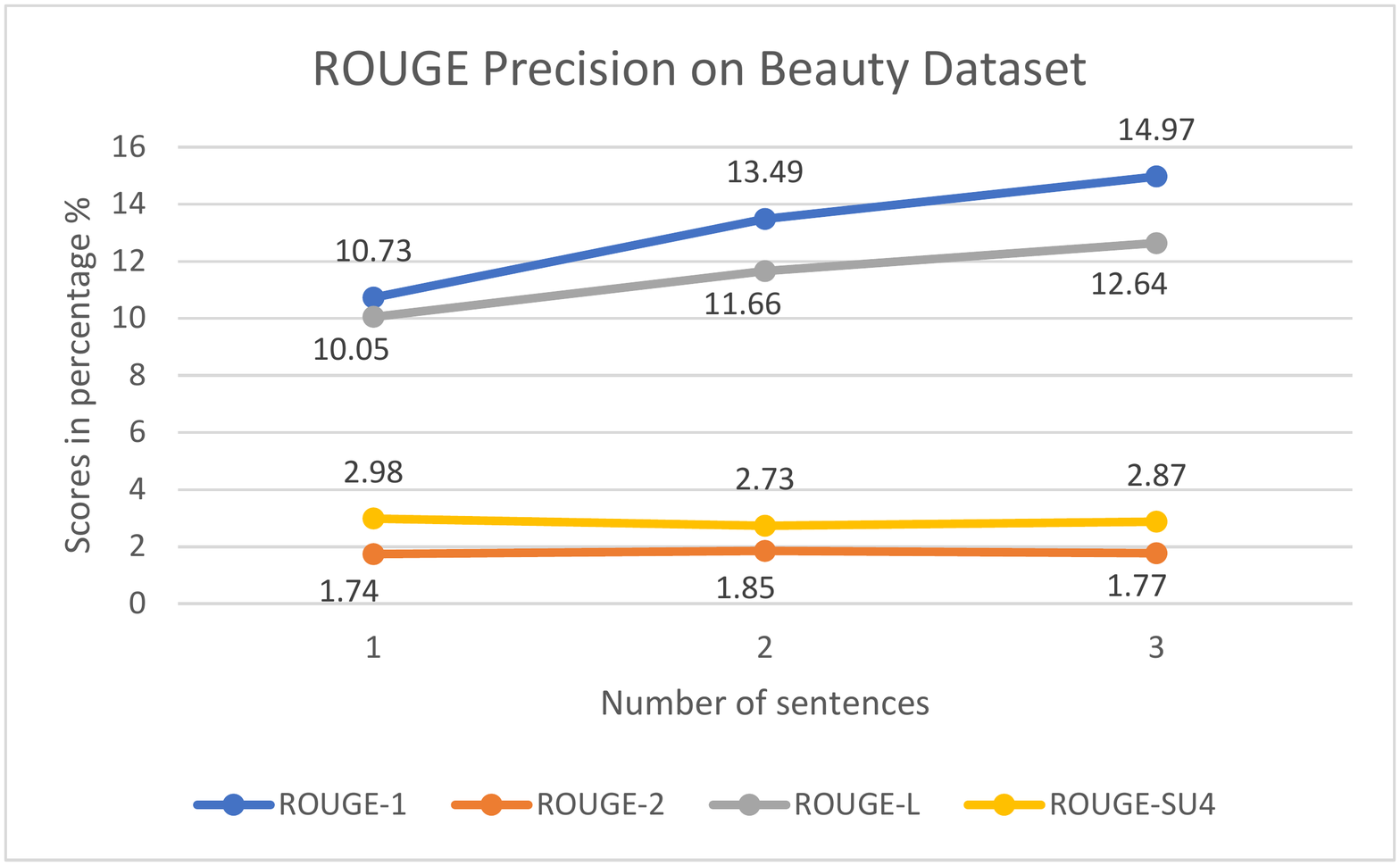}}
\subfigure{\label{fig:rouge_f1}\includegraphics[width=.31\linewidth]{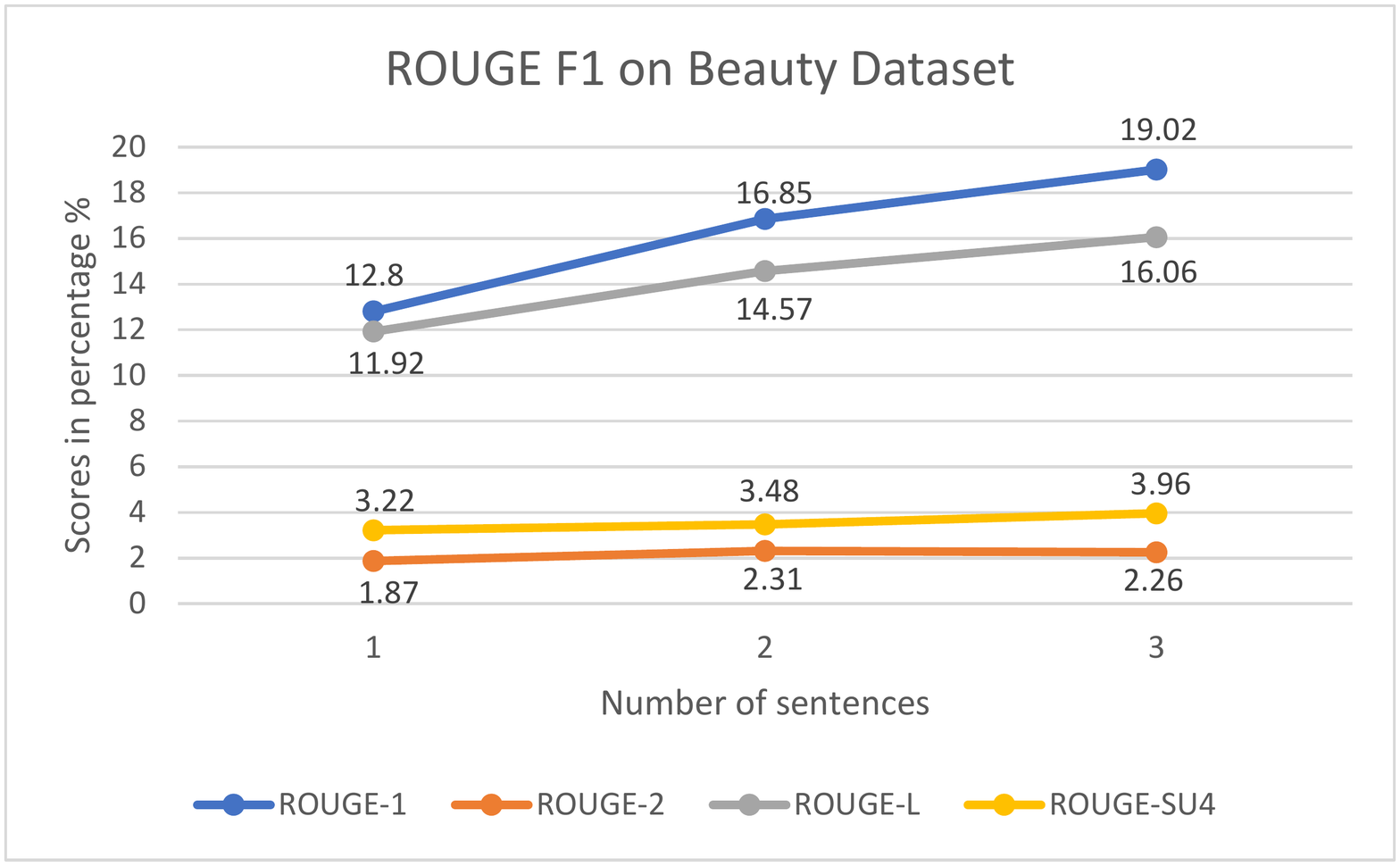}}
\vspace{-15pt}
\end{figure*}

Our model has the ability of generating multiple sentences. To evaluate the multiple sentences generation quality, we do experiments on Beauty dataset. We choose the number of sentences in the range of 1 to 3 during trainig and testing stages. For example, when the number of sentences is set to 1, we only use the first sentence of each review to train the model. During the testing stage, we only generate one sentence and calculate ROUGE and BLEU score based on the first sentence in the human reference text. We report the changing of recall, precision and F-measure of ROUGE scores with respect to the number of sentences in Figure \ref{fig:rouge_recall}, \ref{fig:rouge_precision} and \ref{fig:rouge_f1}. From the results, we can see that our model can have better recall on all the measures when to generate more than one sentence. The ROUGE precision score on multiple sentences generation is slightly lower than the one sentence case. A possible reason is that the more sentences involved in the training and testing, the more challenging for the generation model to cover the information in the human reference texts.

\subsection{Case study}

One thing we need to claim is that we do not do the length alignment on the review data. That means some review only contains one sentence while some of them may contains 2 or more sentences. For each sentence, the length also varies. This is a big challenge to the RNN-based sentence generation model. One reason is that the training of quite long sentences would suffer from gradient vanishing problem which would be hard for deep neural network to learning the parameters. Our hierarchical GRU model could help to solve this problem. It is because that our context level GRU could capture the long dependency so that the length of sequence for each generation process is reduced. The experiment results verify that our model has the ability of generating multiple sentences.

In Table \ref{t:sentence} we list some generated explanations which cover the good sentences with explainability, sentence with feature word but not quite fluent, bad sentence with wrong description of the item. For the last example, the wrong description means that the item is a wireless router but the sentence is not describing the item correctly. This is a common issue we encountered during the experiments. A possible reason is that the dataset is very sparse so the corresponding item vector is not well trained, which result in the wrong description issue. 

\section{Conclusions and Future Work}\label{sec:future}
In this work, we proposed a deep learning framework called hierarchical sequence-to-sequence (HSS) model, which can not only make accurate rating predictions but also generate explanation sentences to improve the effectiveness and the trustworthiness of the recommender systems. For rating prediction, our model can outperform the CF-based BiasedMF and SVD++ algorithms and achieve a comparable result to the state-of-art DeepCoNN model. For the explanation generation module, we designed a hierarchical GRU with feature-aware attention mechanism to generate personalized explanation sentences. We also introduced an auto-denoising method to reduce the effect of unrelated sentences in training process. In the future, we expect to do research work to solve the wrong description issue mentioned in the previous section. We will also apply this framework on other datasets to test its robustness, and consider knowledge-enhanced explanation generation for explainable AI.

\section*{Acknowledgement}
We thank the reviewers for the careful reviews and constructive suggestions. This work was partly supported by the National Science Foundation under IIS-1910154. Any opinions, findings, conclusions or recommendations expressed in this paper are the authors and do not necessarily reflect those of the sponsors.

%\section*{Appendix}

\bibliographystyle{ACM-Reference-Format}
%\balance
\bibliography{paper}

\end{document}